# Double-valued strong-coupling corrections to Bardeen-Cooper-Schrieffer ratios

E. F. Talantsev[1,2]

[1]M.N. Mikheev Institute of Metal Physics, Ural Branch, Russian Academy of Sciences, 18, S. Kovalevskoy St., Ekaterinburg, 620108, Russia

[2]NANOTECH Centre, Ural Federal University, 19 Mira St., Ekaterinburg, 620002, Russia

E-mail: evgeny.talantsev@imp.uran.ru

*Abstract*

Experimental discovery of near-room-temperature (NRT) superconductivity in highly-compressed $H_3S$, $LaH_{10}$ and $YH_6$ restores fundamental interest to electron-phonon pairing mechanism in superconductors. One of prerequisites of phonon-mediated NRT superconductivity in highly-compressed hydrides is strong electron-phonon interaction, which can be quantified by dimensionless ratios of Bardeen-Cooper-Schrieffer (BCS) theory vs $(k_B T_c)/(\hbar \omega_{ln})$ variable, where $T_c$ is the critical temperature and $\omega_{ln}$ is the logarithmic phonon frequency (Mitrovic *et al.* 1984 Phys. Rev. B 29 184). However, all known strong-coupling correction functions for BCS ratios are applicable for $(k_B T_c)/(\hbar \omega_{ln}) < 0.20$, which is not high enough $(k_B T_c)/(\hbar \omega_{ln})$ range for NRT superconductors, because the latter exhibit variable values of $0.13 < (k_B T_c)/(\hbar \omega_{ln}) < 0.32$. In this paper, we reanalyze full experimental dataset (including data for highly-compressed $H_3S$) and find that strong-coupling correction functions for the gap-to-critical-temperature ratio and for the specific-heat-jump ratio are double-valued nearly-linear functions of $(k_B T_c)/(\hbar \omega_{ln})$.

**Double-valued strong-coupling corrections to Bardeen-Cooper-Schrieffer ratios**

I. Introduction

Experimental discovery of near-room-temperature (NRT) superconductivity in highly-compressed $H_3S$ by Drozdov *et al* [1] heralded a new era in condensed matter physics. To date, record transition temperature of $T_c = 260$ K stands with highly-compressed lanthanum superhydride, $LaH_{10}$ [2,3], while recently discovered [4,5] highly-compressed yttrium superhydride, $YH_6$, exhibits $T_c = 240$ K.

Current theoretical understanding of NRT superconductivity in highly-compressed super-hydrides/deuterides is based on a concept of strong electron-phonon pairing [4,6-11] within Eliashberg's theory of superconductivity [12-14]. Due to electron-phonon spectral function, $\alpha^2(\omega) \cdot F(\omega)$, where $F(\omega)$ is the phonon density of states, can be very accurately computed by modern first-principles calculations [4,6-11,15], so-called logarithmic phonon frequency, $\omega_{ln}$, [16,17]:

$$\omega_{ln} = exp\left(\frac{\int_0^\infty \frac{ln(\omega)}{\omega} \cdot F(\omega) \cdot d\omega}{\int_0^\infty \frac{1}{\omega} \cdot F(\omega) \cdot d\omega}\right). \qquad (1)$$

can be also very accurately calculated. Characteristic frequency, $\omega_{ln}$, is used in dimensionless ratio of:

$$\frac{k_B \cdot T_c}{\hbar \cdot \omega_{ln}} \qquad (2)$$

where $k_B$ is the Boltzmann constant, and $\hbar$ is reduced Planck constant. This ratio is often referred in its reduced form of $\frac{T_c}{\omega_{ln}}$.

Mitrovic *et al* [18] proposed to use $\frac{k_B \cdot T_c}{\hbar \cdot \omega_{ln}}$ as a primary variable in the strong-coupling correction function for gap-to-critical-temperature ratio, $\frac{2 \cdot \Delta(0)}{k_B \cdot T_c}$. Later, Marsiglio and Carbotte [19], and Carbotte [20] extended this proposal and used $\frac{k_B \cdot T_c}{\hbar \cdot \omega_{ln}}$ as a variable in strong-coupling correction functions for other dimensionless ratios of Bardeen-Cooper-Schrieffer (BCS) theory [21]. All proposed correction functions [18-20] have general form:

$$\begin{aligned}\frac{2 \cdot \Delta(0)}{k_B \cdot T_c} &= \\ \frac{\Delta C(T_c)}{\gamma \cdot T_c} &= ln\left(\left(e \cdot \left(\frac{1}{C} \cdot \frac{\hbar \cdot \omega_{ln}}{k_B \cdot T_c}\right)^{-B \cdot \left(\frac{k_B \cdot T_c}{\hbar \cdot \omega_{ln}}\right)^2}\right)^A\right) \\ \frac{\mu_0^2 \cdot \gamma \cdot T_c}{B_c^2(0)} &= \end{aligned} \quad (3)$$

where A, B, and C are fitting constants, $\Delta(0)$ is the amplitude of the ground state energy gap, $\Delta C(T_c)$ is specific heat jump at $T_c$, γ is Sommerfeld constant, μ$_0$ is the permeability of free space, and $B_c = \frac{\phi_0}{2 \cdot \sqrt{2} \cdot \pi} \cdot \frac{1}{\lambda \cdot \xi}$ is thermodynamic critical field, where ϕ$_0$ is flux quantum, λ is London penetration depth, and ξ is the coherence length.

Eq. 3 is in a wide use in studies of highly-compressed hydrides [4,6-11], due to there is general consensus between theoretical groups [22], that NRT superconductivity is originated from strong electron-phonon interaction. There is also another consensus [4-11,19,20] that Eq. 3 accurately describes experimental data for:

$$\frac{k_B \cdot T_c}{\hbar \cdot \omega_{ln}} \leq 0.25. \quad (4)$$

II. Description of the problem

It should be noted that Eq. 3 is based on non-linear fitting term proposed by Geilikman and Kresin [23]:

$$f(x) = ln\left(\left(\frac{1}{b \cdot x}\right)^{a \cdot x^2}\right) \quad (5)$$

where *a* and *b* are free fitting parameters, and $x = \frac{k_B \cdot T_c}{\hbar \cdot \omega_0}$, where ω$_0$ is some characteristic frequency of full phonon spectrum, for which Mitrovic *et al* [18] proposed to use ω$_{ln}$ (Eq. 1).

It can be seen that Eq. 5 and, as a consequence, Eq. 3 have one hidden parameter, which is power law exponent 2, and general formula for Eq. 5 is:

$$f(x) = a \cdot x^c \cdot ln\left(\frac{1}{b \cdot x}\right) = ln\left(\left(\frac{1}{b \cdot x}\right)^{a \cdot x^c}\right) \tag{6}$$

where, $a$, $b$, and $c$ are free fitting parameters, while Eq. 3 should be expressed:

$$\begin{aligned}\frac{2 \cdot \Delta(0)}{k_B \cdot T_c} &= \\ \frac{\Delta C(T_c)}{\gamma \cdot T_c} &= ln\left(\left(e \cdot \left(\frac{1}{C} \cdot \frac{\hbar \cdot \omega_{ln}}{k_B \cdot T_c}\right)^{-B \cdot \left(\frac{k_B \cdot T_c}{\hbar \cdot \omega_{ln}}\right)^D}\right)^A\right) \\ \frac{\mu_0^2 \cdot \gamma \cdot T_c}{B_c^2(0)} &= \end{aligned} \tag{7}$$

where A, B, C, and D are free-fitting parameters. However, Eq. 7 even in its reduced form, i.e. Eq. 3, cannot have completely independent parameters, because of its complexity and strong non-linearity.

Despite mentioned above consensus about strong-coupling nature of electron-phonon interaction in NRT superconductors, analyses of experimental self-field critical current, $J_c(sf,T)$, data [24] and the upper critical field, $B_{c2}(T)$, data [25] in highly-compressed $H_3S$ showed that:

$$\frac{2 \cdot \Delta(0)}{k_B \cdot T_c} = 3.20 \pm 0.03 \qquad \text{(Ref. 24)} \tag{8}$$

$$\frac{2 \cdot \Delta(0)}{k_B \cdot T_c} = 3.55 \pm 0.31 \qquad \text{(Ref. 25)} \tag{9}$$

$$\frac{\Delta C(T_c)}{\gamma \cdot T_c} = 1.33 \pm 0.25 \qquad \text{(Ref. 25)} \tag{10}$$

which all are within weak-coupling limits of the BCS theory (which are $\frac{2 \cdot \Delta(0)}{k_B \cdot T_c} = 3.53$ and $\frac{\Delta C(T_c)}{\gamma \cdot T_c} = 1.43$), while first principles calculations [26-28] showed that:

$$4.4 \leq \frac{2 \cdot \Delta(0)}{k_B \cdot T_c} \leq 4.9 \tag{11}$$

$$\frac{\Delta C(T_c)}{\gamma \cdot T_c} \cong 2.7 \tag{12}$$

In regard of lower than weak-coupling value for the $\frac{2 \cdot \Delta(0)}{k_B \cdot T_c}$ ratio deduced in Ref. 24 (Eq. 8), we can refer early experimental results for pure aluminium (which is considered as canonical weak-coupling BCS superconductor):

$$\frac{2 \cdot \Delta(0)}{k_B \cdot T_c} = 3.15 \pm 0.22 \qquad (13)$$

reported by Douglass and Meservey [29] in 1964. In this regard, the value of $\frac{2 \cdot \Delta(0)}{k_B \cdot T_c} = 3.20 \pm 0.03$ [24] has been deduced for a single dataset (available at that time and this is still the only $J_c(\text{sf},T)$ dataset available to date) which has six data points and most of these data points were measured at high reduced temperatures. However, the analysis [24] unavoidably showed that H₃S is definitely not strong-coupling superconductor, because $J_c(\text{sf},T)$ data fit at fixed ratio of $\frac{2 \cdot \Delta(0)}{k_B \cdot T_c} = 4.5$ completely differs from experimental data (see Fig. 1 in Ref. 24).

The value of $\frac{2 \cdot \Delta(0)}{k_B \cdot T_c} = 3.55 \pm 0.31$ [25] deduced from the analysis of four $B_{c2}(T)$ datasets reported by Mozaffari *et al.* [30], each of those has at least 15 experimental data points covered significant part of full temperature range of 0 K < $T/T_c$ < $T_c$. As the result, deduced value for $\frac{2 \cdot \Delta(0)}{k_B \cdot T_c}$ is within weak-coupling phenomenology.

Despite a fact that Eq. 3 is in a wide used to study superconductors ranging from elements [20,30] to NRT superconductors [4,7-9], it should be stressed that Mitrovic *et al.* [18], Marsiglio and Carbotte [19], Carbotee [20], and Nicol and Carbotee [7] in their analyses excluded data points with ratios of $\frac{k_B \cdot T_c}{\hbar \cdot \omega_{ln}} > 0.20$. However, these data points are presented in Table IV and Table X in Ref. 20. We highlight excluded data points in $\frac{2 \cdot \Delta(0)}{k_B \cdot T_c}$ vs $\frac{k_B \cdot T_c}{\hbar \cdot \omega_{ln}}$ plot by red circuit (Fig. 1).

However, recent first-principles calculations reported by Troyan *et al.* [4] showed that $Im\bar{3}m$-YH₆ phase at $P$ = 120-200 GPa (with experimentally observed $T_c$ = 220-240 K [4,5]) exhibits

$$0.22 \leq \frac{k_B \cdot T_c}{\hbar \cdot \omega_{ln}} \leq 0.23 \qquad (14)$$

for which respectful value deduced from extrapolative curve proposed by Mitrovic *et al.* [18] is:

$$4.9 \leq \frac{2\cdot\Delta(0)}{k_B\cdot T_c} \leq 5.5 \qquad (15)$$

However, an extrapolation of Eq. 3 in the region of $\frac{k_B\cdot T_c}{\hbar\cdot\omega_{ln}} > 0.20$ (with A, B, and C parameters reported by Mitrovic *et al* [18]) cannot be accurate, because data for $\frac{k_B\cdot T_c}{\hbar\cdot\omega_{ln}} > 0.20$ was excluded in Ref. 18 from the consideration. Fig. 1 also shows that there is no electron-phonon mediated material which exhibits $\frac{2\cdot\Delta(0)}{k_B\cdot T_c} \geq 5.2$.

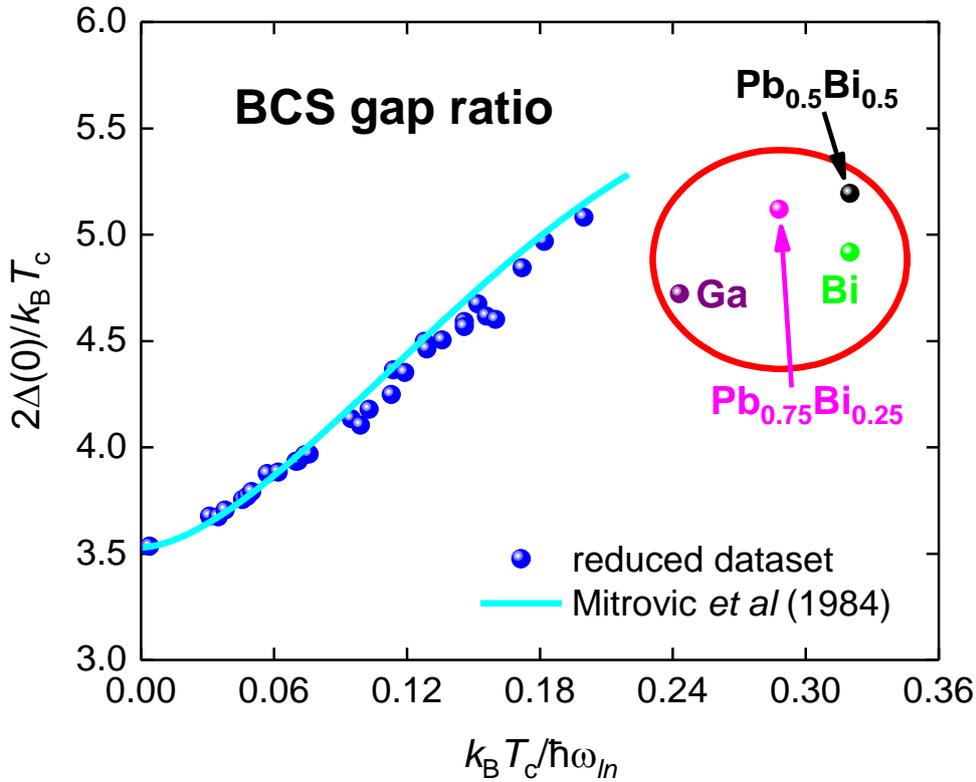

**Figure 1.** The gap ratio $\frac{2\cdot\Delta(0)}{k_B\cdot T_c}$ vs $\frac{k_B\cdot T_c}{\hbar\cdot\omega_{ln}}$. The data points are taken from Table IV of Ref. 20. The cyan line is Eq. 3 proposed by Mitrovic *et al* [18]. Blue points are the data used by Mitrovic *et al* [18]. Data points in red circle were excluded from the consideration in Refs. [7,18-20].

Taking in account that recently Kruglov *et al.* [8] calculate that $Fm\overline{3}m$ phase of LaH$_{10}$ at $P$ = 170-210 GPa exhibits

$$0.29 \leq \frac{k_B\cdot T_c}{\hbar\cdot\omega_{ln}} \leq 0.32 \qquad (16)$$

with respectful value of

$$5.45 \leq \frac{2 \cdot \Delta(0)}{k_B \cdot T_c} \leq 5.55 \tag{17}$$

there is a need to reconsider strong-coupling correction functions for BCS ratios for the case of $\frac{k_B \cdot T_c}{\hbar \cdot \omega_{ln}} > 0.20$. This analysis is presented herein.

There is a need to make a clarity, that an approach to restrict total databased before the analysis has general designation as survivorship bias [31,32], when a portion of full experimental database is excluded from the consideration by applying some hidden or clearly stated criterion (in given case, the hidden rule was $\frac{k_B \cdot T_c}{\hbar \cdot \omega_{ln}} \leq 0.20$), while deduced function/dependence is extrapolated on a sampling range which has been excluded from the consideration (which is in given case, $\frac{k_B \cdot T_c}{\hbar \cdot \omega_{ln}} > 0.20$).

**III. Results and Discussions**

**3.1. Exact formula for $\frac{2 \cdot \Delta(0)}{k_B \cdot T_c}$ vs $\frac{k_B \cdot T_c}{\hbar \cdot \omega_{ln}}$ within Mitrovic's approach**

We test Eq. 3 to satisfy the criteria of the accuracy and of mutual parameters independence by fitting $\frac{2 \cdot \Delta(0)}{k_B \cdot T_c}$ data of the restricted dataset (blue data points in Fig. 1, which are taken from Table IV of Ref. 20) which was used by Mitrovic *et al.* [18] in their derivation of the equation [18]:

$$\frac{2 \cdot \Delta(0)}{k_B \cdot T_c} = ln\left(\left(e \cdot \left(\frac{1}{2} \cdot \frac{\hbar \cdot \omega_{ln}}{k_B \cdot T_c}\right)^{12.5 \cdot \left(\frac{k_B \cdot T_c}{\hbar \cdot \omega_{ln}}\right)^2}\right)^{3.53}\right) \tag{18}$$

In our fits, data points exhibited high $\frac{k_B \cdot T_c}{\hbar \cdot \omega_{ln}}$ values (which were excluded from the analysis in Refs. 7,18-20) were not fitted too (but ones are shown in Fig. 2), to make a direct comparison numerical constants reported by Mitrovic *et al.* [18] (Eq. 18) with free-fitting parameters of Eq. 3. Results of fits are summarized in Table I and some fits are shown in Figure 2.

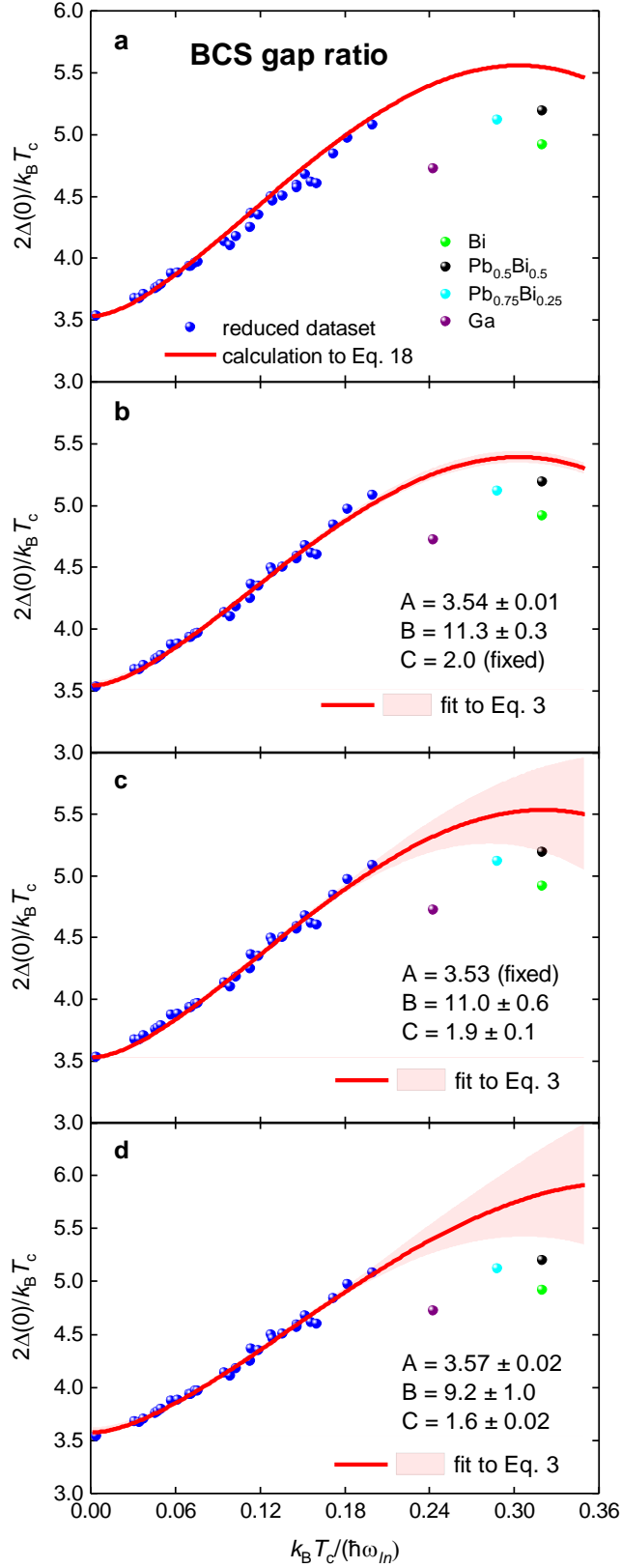

**Figure 2.** The gap ratio $\frac{2 \cdot \Delta(0)}{k_B \cdot T_c}$ vs $\frac{k_B \cdot T_c}{\hbar \cdot \omega_{ln}}$. Data is taken from Table IV of Ref. 20. Blue data points are used for fits (b-d). 95% confidence bands are shown. a – calculation to Eq. 18; b – fit to Eq. 3, goodness of fit $R = 0.990$; c – fit to Eq. 3, $R = 0.990$; d – fit to Eq. 3, $R = 0.992$.

It can be seen (Table 1 and Figure 2) that Eq. 14 does not provide a good quality for the dataset ($R = 0.965$), and what is also important, that our fits to Eq. 3 reveal completely different A, B, and C parameters values (Table 1 and Fig. 2):

$$\frac{2 \cdot \Delta(0)}{k_B \cdot T_C} = ln\left(\left(e \cdot \left(\frac{1}{1.6} \cdot \frac{\hbar \cdot \omega_{ln}}{k_B \cdot T_C}\right)^{9.2 \cdot \left(\frac{k_B \cdot T_C}{\hbar \cdot \omega_{ln}}\right)^2}\right)^{3.57}\right) \qquad (19)$$

Taking in account that fitting function (Eq. 3) is strongly non-linear, Eq. 19 is remarkably different from Eq. 18, including the power exponent of 3.57 (Eq. 19) vs expected and reported [18] value of 3.53 (Eq. 18).

**Table 1.** Results of fits of restricted $\frac{2 \cdot \Delta(0)}{k_B \cdot T_C}$ dataset [7,18-20] to Eqs. 3,18. Fixed parameters are indicated in bold red.

| Model | A | B | C | Goodness of fit, $R$ | Mutual parameters dependence |
|---|---|---|---|---|---|
| Eq. 14 | **3.53** | **12.5** | **2.0** | 0.965 | Not applicable |
| Eq. 3 | **3.53** | **12.5** | 2.18 ± 0.03 | 0.988 | Not applicable |
|  | **3.53** | 11.6 ± 0.1 | **2.0** | 0.989 | Not applicable |
|  | 3.48 ± 0.01 | **12.5** | **2.0** | 0.983 | Not applicable |
|  | 3.54 ± 0.01 | 11.3 ± 0.3 | **2.0** | 0.990 | 0.804 |
|  | 3.52 ± 0.01 | **12.5** | 2.15 ± 0.04 | 0.988 | 0.679 |
|  | **3.53** | 11.0 ± 0.6 | 1.9 ± 0.1 | 0.990 | 0.968 |
|  | 3.57 ± 0.02 | 9.2 ± 1.0 | 1.6 ± 0.2 | 0.992 | 0.898-0.992 |
|  | **3.57** | **9.2** | **1.6** | 0.992 | Not applicable |

### 3.2. Equation for the gap-to-critical-temperature ratio

The large difference between Eq. 18 and Eq. 19 reflects a simple fact that restricted experimental dataset (blue data points in Figs. 1,2) is practically a linear function. By employing strongly non-linear fitting function, i.e. Eq. 3 [7,18-20], to fit nearly linear dependent dataset causes a problem which is known as an overfitting problem. Truly, it can be seen in Table 1, that for the case of three free fitting parameters, there are large mutual parameters dependences. General solution for overfitting problem is to find a simple function with minimal number of free fitting parameters which fits the data with similar quality.

By considering several possible options, we find that the restricted dataset can to be fitted to simple function of:

$$\frac{2\cdot\Delta(0)}{k_B\cdot T_c} = A \cdot \left(1 + B \cdot \left(\frac{k_B \cdot T_c}{\hbar \cdot \omega_{ln}}\right)^C\right). \qquad (20)$$

Fitting results are summarized in Table 2 and showed in Fig. 3.

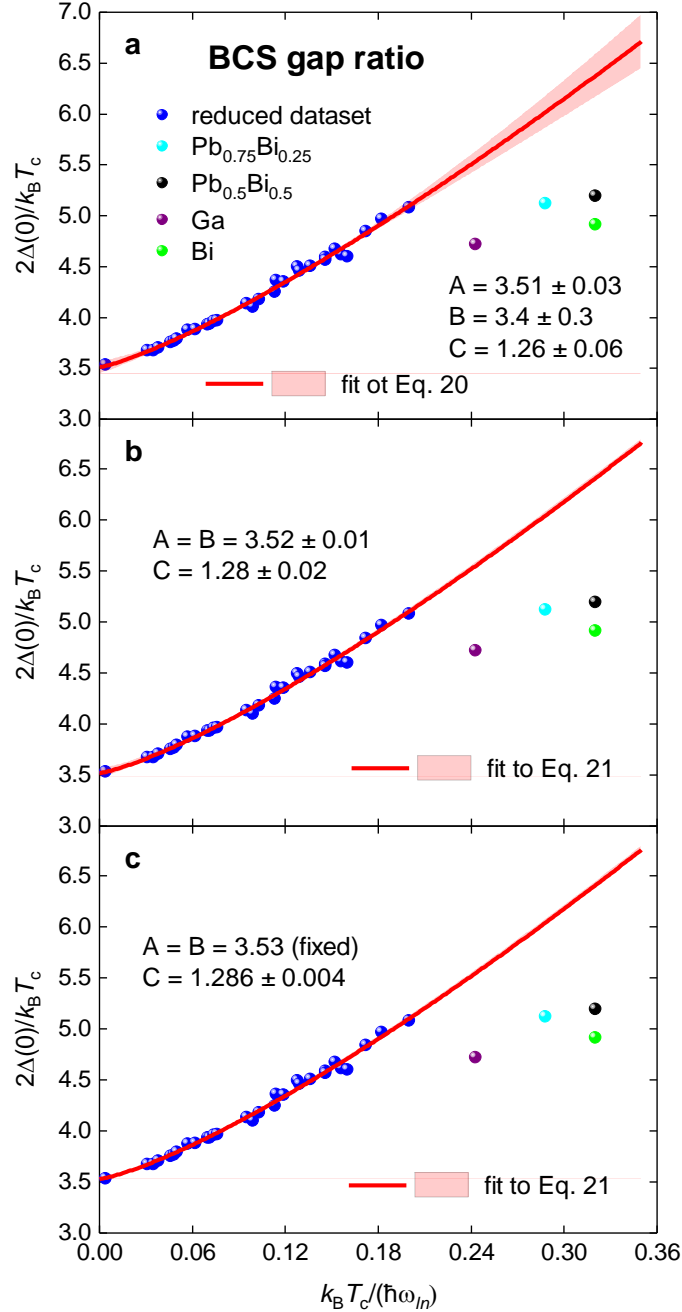

**Figure 3.** The gap ratio $\frac{2\cdot\Delta(0)}{k_B\cdot T_c}$ vs $\frac{k_B\cdot T_c}{\hbar\cdot\omega_{ln}}$ and data fits to Eqs. 20,21. Data is taken from Table IV of Ref. 21. Blue data points are used for fits. 95% confidence bands are shown. a - three free-fitting parameters, $R = 0.992$; b - A = B, $R = 0.992$; c - A = B = 3.53, $R = 0.992$.

It can be seen (Table 2, Fig. 3) that free-fitting parameters of A and B are close to each other and simultaneously close to weak-coupling limit of 3.53. Based on this, further reduction in number of parameters has be done by applying the condition of A = B:

$$\frac{2 \cdot \Delta(0)}{k_B \cdot T_c} = A \cdot \left(1 + A \cdot \left(\frac{k_B \cdot T_c}{\hbar \cdot \omega_{ln}}\right)^C\right) \qquad (21)$$

The fit quality does not change ($R = 0.992$) by applying this condition and, in addition, we observed substantial drop in mutual parameters dependence to 0.938.

Based on a fact that free-fitting parameter $A = B = 3.52 \pm 0.01$ is remarkably close to the BCS weak-coupling limit of 3.53, Eq. 21 is further simplified by applying condition of A = B = 3.53 and final equation is simplified to unexpected elegant form:

$$\frac{2 \cdot \Delta(0)}{k_B \cdot T_c} = 3.53 \cdot \left(1 + 3.53 \cdot \left(\frac{k_B \cdot T_c}{\hbar \cdot \omega_{ln}}\right)^{1.29}\right) \qquad (22)$$

Despite a fact that at the moment there is no interpretation for power-law exponent of 1.29 in Eq. 22, general form of Eq. 22 satisfies to at least two signs of the correctness of physical law: the simplicity and the charm (to prove this, one can make comparison of Eq. 22 with Eqs. 3,18,19).

**Table 2.** Results of fits of restricted $\frac{2 \cdot \Delta(0)}{k_B \cdot T_c}$ dataset [7,18-20] to Eq. 20. Fixed parameters are indicated in bold red.

| Model | A | B | C | Goodness of fit, $R$ | Mutual parameters dependence |
|---|---|---|---|---|---|
| Eq. 16 | $3.51 \pm 0.03$ | $3.4 \pm 0.3$ | $1.26 \pm 0.06$ | 0.992 | 0.962-0.995 |
|  | **A = B** $3.52 \pm 0.01$ | **A = B** $3.52 \pm 0.01$ | $1.28 \pm 0.02$ | 0.992 | 0.938 |
|  | **3.53** | **3.53** | $1.286 \pm 0.004$ | 0.992 | Not applicable |

### 3.3. Double-valued $\frac{2 \cdot \Delta(0)}{k_B \cdot T_c}$ vs $\frac{k_B \cdot T_c}{\hbar \cdot \omega_{ln}}$ correction function

First principles calculations show that highly-compressed $H_3S$ has $\frac{k_B \cdot T_c}{\hbar \cdot \omega_{ln}} = 0.134$ [26]. Based on this result, we add $H_3S$ in $\frac{2 \cdot \Delta(0)}{k_B \cdot T_c}$ vs $\frac{k_B \cdot T_c}{\hbar \cdot \omega_{ln}}$ plot (Fig. 4), where we use $\frac{2 \cdot \Delta(0)}{k_B \cdot T_c} = 3.55$, which was deduced in our previous work [25] from temperature dependent $B_{c2}(T)$ data reported by Mozafari *et al.* [33]. It can be seen that H$_3$S (red data point in Fig. 4) is located well below a trendline where the majority of superconductors are located.

However, it can be also seen in Fig. 4, that H$_3$S and superconductors with high $\frac{k_B \cdot T_c}{\hbar \cdot \omega_{ln}}$ ratios (which were excluded in previous considerations from the analysis [7,18-20]) can potentially form the second trendline which located below the trendline described by Eq. 20. To be consistent with our approach developed above (and very limited experimental data for this second trendline), we fit this dataset (of high $\frac{k_B \cdot T_c}{\hbar \cdot \omega_{ln}}$ ratios and H$_3$S) to fitting equation with one free-fitting parameter, which is the amplitude, and variable part of fitting equation was fixed to be the same as in Eq. 22:

$$\frac{2 \cdot \Delta(0)}{k_B \cdot T_c} = A \cdot \left(1 + 3.53 \cdot \left(\frac{k_B \cdot T_c}{\hbar \cdot \omega_{ln}}\right)^{1.29}\right) \qquad (23)$$

Fitting result to Eq. 23 is shown in Fig. 4, where free-fitting parameter $A = 2.87 \pm 0.06$. It is important to note, that 95% confidence band for this fit covers all data in the dataset.

Based on this, we can conclude that strong-correction function for the gap-to-critical-temperature ratio is double-valued function. The existence of the second branch explains the contradiction between the first-principles calculations, which showed that H$_3$S has the ratio of $\frac{k_B \cdot T_c}{\hbar \cdot \omega_{ln}} = 0.134$ and based on conventional (the upper) trendline, this compound should have $\frac{2 \cdot \Delta(0)}{k_B \cdot T_c} \gtrsim 4.5$, and the analysis of experimental data [24,25] which showed that H$_3$S is weak-coupling superconductor.

Because all first principles calculations [2,4,6,7,9-11,22,26-28,34] show that NRT superconductors have a large or very large $\frac{k_B \cdot T_c}{\hbar \cdot \omega_{ln}}$ value and, thus, based on the upper branch of the $\frac{2 \cdot \Delta(0)}{k_B \cdot T_c}$ correction function these superconductors are forced to have extremely high gap-to-critical-temperature ratio $\frac{2 \cdot \Delta(0)}{k_B \cdot T_c} \gtrsim 5.0$, the existence of the lower branch removes this enforcement, and despite a fact that NRT superconductors might have high $\frac{k_B \cdot T_c}{\hbar \cdot \omega_{ln}}$ variable ones can have moderate values for the gap-to-critical-temperature ratio (Fig. 4).

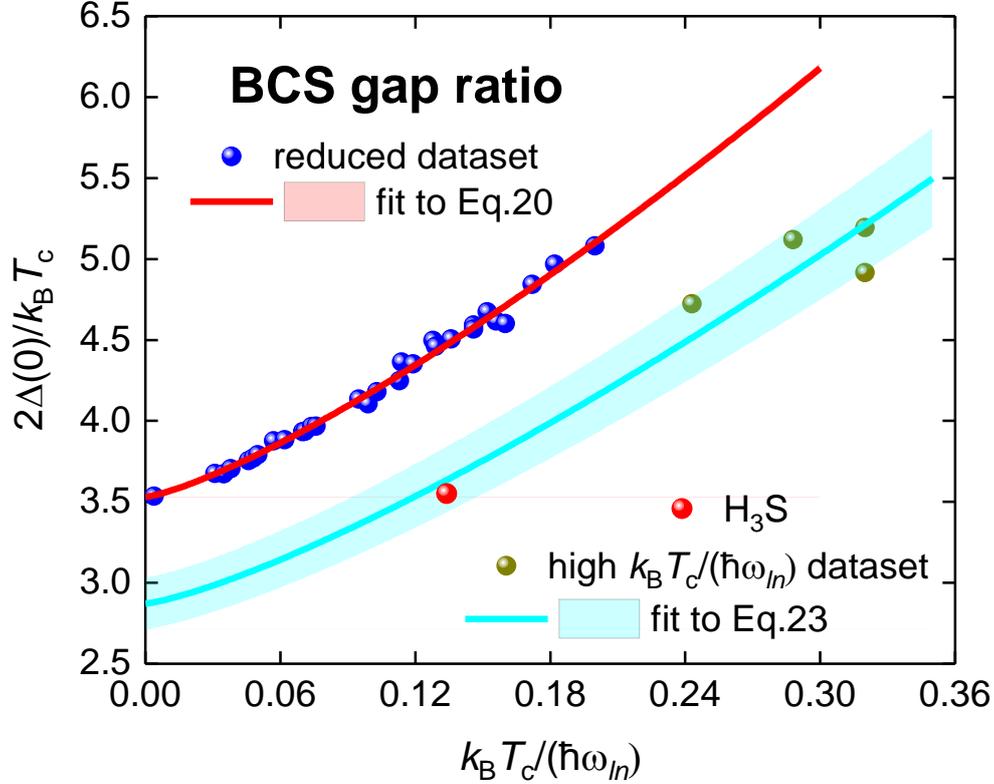

**Figure 4.** Full dataset for the gap ratio $\frac{2 \cdot \Delta(0)}{k_B \cdot T_c}$ vs $\frac{k_B \cdot T_c}{\hbar \cdot \omega_{ln}}$ from Table IV of Ref. 20 and data point for highly-compressed $H_3S$ (red point, see details in main text). Fits to Eq. 20 (blue data points, red curve) and Eq. 23 (dark yellow data points and red point, cyan curve) are shown together with 95% confidence bands. The lower curve (Eq. 23) has free-fitting parameter $A = 2.87 \pm 0.06$ and $R = 0.899$.

### 3.4. Exact formula for $\frac{\Delta C(T_c)}{\gamma \cdot T_c}$ vs $\frac{k_B \cdot T_c}{\hbar \cdot \omega_{ln}}$ within Marsiglio-Carbotte approach

Specific-heat-jump ratio at the transition temperature, $\frac{\Delta C(T_c)}{\gamma \cdot T_c}$, is another widely used dimensionless ratio within electron-phonon mediated phenomenology for which strong coupling correction function (Eq. 3) was proposed by Marsiglio and Carbotte [19]. In Fig. 5 we show $\frac{\Delta C(T_c)}{\gamma \cdot T_c}$ vs $\frac{k_B \cdot T_c}{\hbar \cdot \omega_{ln}}$ dataset from Table X of Ref. 20. Superconductors exhibited $\frac{k_B \cdot T_c}{\hbar \cdot \omega_{ln}} \geq$ 0.20 were not taken in the fit in Fig. 53 of Ref. 20, and we highlight these superconductors by read oval in Fig. 5. We also add data for highly-compressed $H_3S$ (Eq. 20), for which we used calculated $\frac{k_B \cdot T_c}{\hbar \cdot \omega_{ln}} = 0.134$ [26] and experimental value of $\frac{\Delta C(T_c)}{\gamma \cdot T_c} = 1.33$ [25].

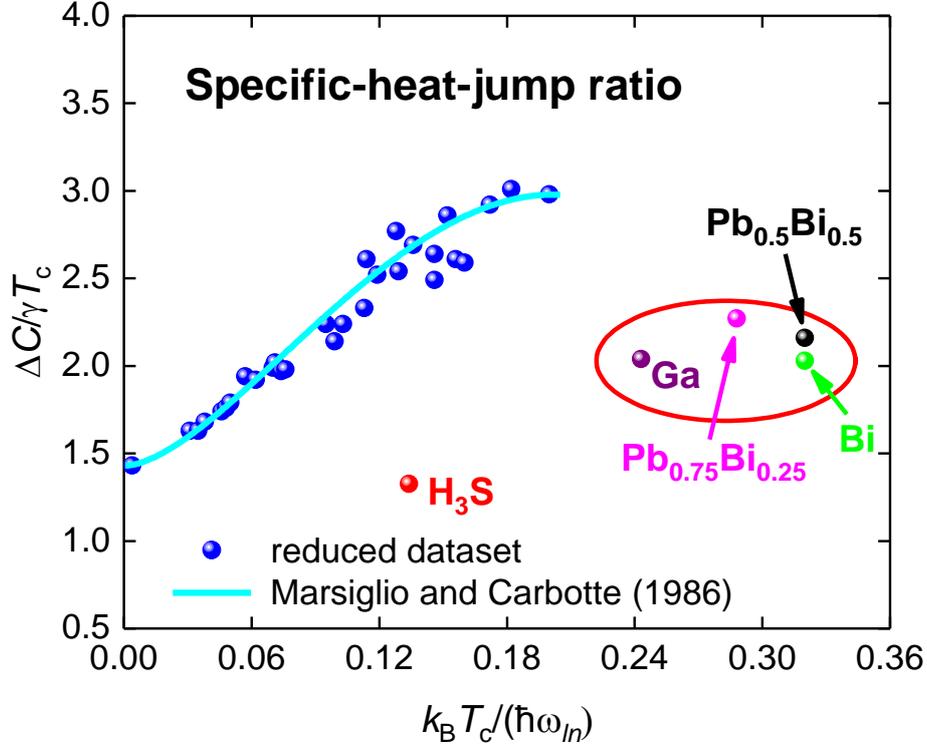

**Figure 5.** Specific-heat-jump ratio $\frac{\Delta C(T_c)}{\gamma \cdot T_c}$ vs $\frac{k_B \cdot T_c}{\hbar \cdot \omega_{ln}}$. The data is taken from Table X of Ref. 21 and data for $H_3S$ is taken from Refs. 25,26. The cyan line is Eq. 3 with A = 1.43, B = 52, C = 3. Data points in red oval were excluded from consideration by Marsiglio and Carbotte [19] and Carbotte [20] in their derivation of parameters for Eq. 3.

It can be seen (Fig. 5) that even restricted $\frac{\Delta C(T_c)}{\gamma \cdot T_c}$ vs $\frac{k_B \cdot T_c}{\hbar \cdot \omega_{ln}}$ dataset (blue data points in Fig. 5) has a large scattering, especially for the range of variable within:

$$0.09 < \frac{\Delta C(T_c)}{\gamma \cdot T_c} < 0.18. \qquad (24)$$

An application of non-linear fitting function (Eq. 3):

$$\frac{\Delta C(T_c)}{\gamma \cdot T_c} = ln\left(\left(e \cdot \left(\frac{1}{C} \cdot \frac{\hbar \cdot \omega_{ln}}{k_B \cdot T_c}\right)^{B \cdot \left(\frac{k_B \cdot T_c}{\hbar \cdot \omega_{ln}}\right)^2}\right)^A\right) \qquad (25)$$

with three free-fitting parameters A, B, and C for such scattered dataset cannot be proved to be valid, because the goodness of fit will be always low (for instance, $R = 0.940$ for the fit to Eq. 25), and it cannot be significantly improved because of raw data scattering. Thus, there is a task to find much simpler function which can fit data with similar or better quality.

In Fig. 6(a) we show calculated curve for Eq. 3 for full range of $\frac{k_B \cdot T_c}{\hbar \cdot \omega_{ln}}$, where we used A = 1.43, B = 52, and C = 3 reported by Marsiglio and Carbotte [19]. The curve behaves much better in comparison with its counterpart (i.e., Eq. 18), because it goes down for $\frac{k_B \cdot T_c}{\hbar \cdot \omega_{ln}} > 0.20$ towards to reach experimental data on the high end of $\frac{k_B \cdot T_c}{\hbar \cdot \omega_{ln}}$ values. Nevertheless, our fit of the restricted dataset to Eq. 25 (Fig. 6(b)) reveals that the equation is (Table 3):

$$\frac{\Delta C(T_c)}{\gamma \cdot T_c} = ln\left(\left(e \cdot \left(\frac{\hbar \cdot \omega_{ln}}{2.7 \cdot k_B \cdot T_c}\right)^{42 \cdot \left(\frac{k_B \cdot T_c}{\hbar \cdot \omega_{ln}}\right)^2}\right)^{1.47}\right) \quad (26)$$

which has (similarly to our previous finding in regard of Eq. 18) remarkably different parameters in comparison with ones reported by Marsiglio and Carbotte [19]. More details can be found in Table 3.

**Table 3.** Results of fits of restricted $\frac{\Delta C(T_c)}{\gamma \cdot T_c}$ dataset [19,20] to Eq. 3 and Eq. 25. Fixed parameters are indicated in bold red.

| Model | A | B | C | Goodness of fit, R | Mutual parameters dependence |
|---|---|---|---|---|---|
| Eq. 3 | **1.43** | **53** | **3.0** | 0.940 | Not applicable |
| | **1.43** | **53** | 3.10 ± 0.05 | 0.949 | Not applicable |
| | **3.53** | 50.4 ± 1.0 | **3.0** | 0.952 | Not applicable |
| | 1.40 ± 0.01 | **53** | **3.0** | 0.951 | Not applicable |
| | 1.44 ± 0.04 | 49.7 ± 3.4 | **3.0** | 0.952 | 0.912 |
| | 1.41 ± 0.02 | **53** | 3.03 ± 0.09 | 0.951 | 0.755 |
| | **1.43** | 47.8 ± 3.6 | 2.9 ± 0.2 | 0.953 | 0.930 |
| | 1.47 ± 0.05 | 42 ± 7 | 2.7 ± 0.3 | 0.954 | 0.941-0.985 |
| | **1.47** | **42** | **2.7** | 0.954 | Not applicable |

It can be seen in Fig. 6, that 95% confidence band becomes very wide for $\frac{k_B \cdot T_c}{\hbar \cdot \omega_{ln}} \geq 0.24$, and thus any extrapolations of this equation in regard of NRT superconductors will be unlikely to be convincing.

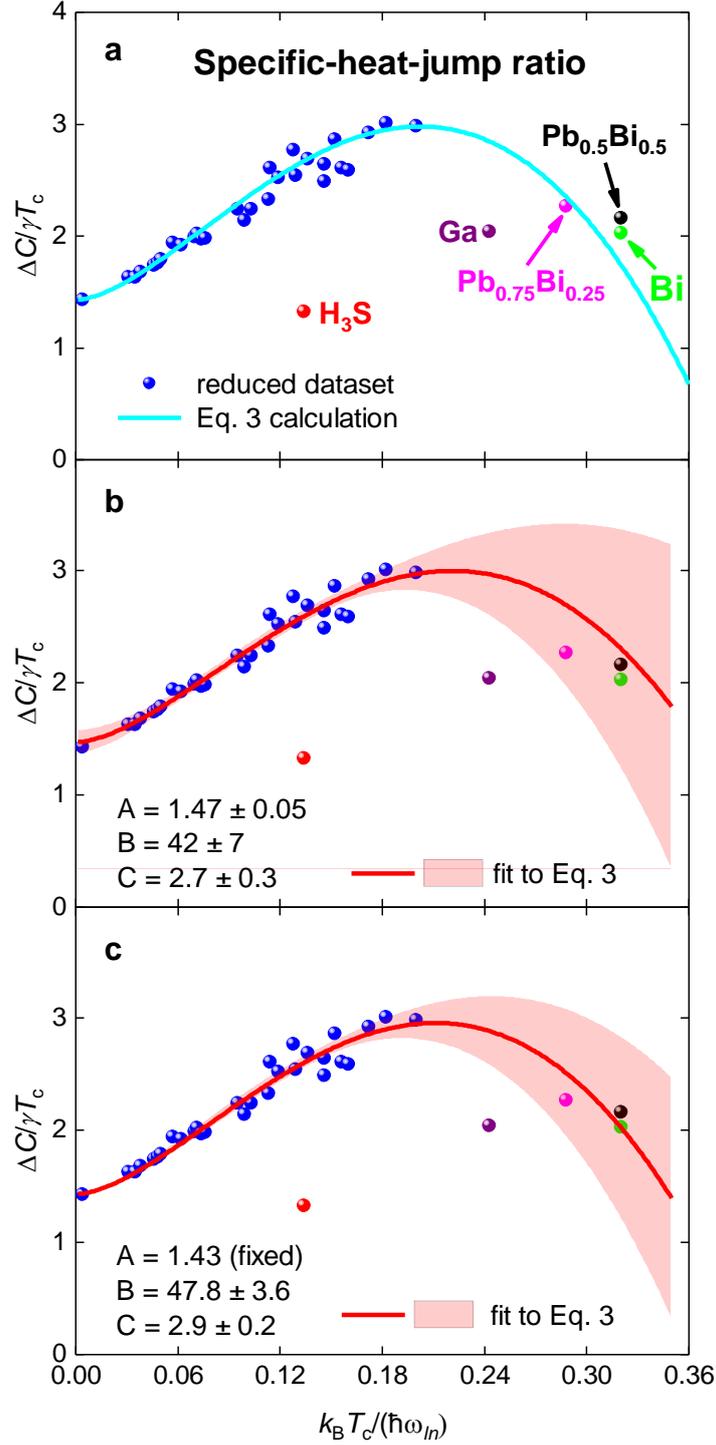

**Figure 6.** Fitting specific-heat-jump ratio $\frac{\Delta C(T_c)}{\gamma \cdot T_c}$ vs $\frac{k_B \cdot T_c}{\hbar \cdot \omega_{ln}}$ fits to Eq. 3. Data is taken from Table X of Ref. 21. Blue data points are used for fits. 95% confidence bands are shown. a – calculation to Eq. 3 for A = 1.43, B = 52, and C = 3 reported by Marsiglio and Carbotte [19]; b – fit to Eq. 3, goodness of fit $R$ = 0.954; c – fit to Eq. 3, $R$ = 0.953.

## 3.5. Double-valued $\frac{\Delta C(T_c)}{\gamma \cdot T_c}$ vs $\frac{k_B \cdot T_c}{\hbar \cdot \omega_{ln}}$ correction function

It can be seen in Fig. 5 that full $\frac{\Delta C(T_c)}{\gamma \cdot T_c}$ vs $\frac{k_B \cdot T_c}{\hbar \cdot \omega_{ln}}$ dataset (including, highly-compressed $H_3S$) can be splatted in two nearly linear-dependent branches. Thus, we fit the restricted dataset indicated by blue data points in Fig. 7 (the upper branch) to linear equation:

$$\frac{\Delta C(T_c)}{\gamma \cdot T_c} = A \cdot \left(1 + B \cdot \left(\frac{k_B \cdot T_c}{\hbar \cdot \omega_{ln}}\right)\right) \qquad (27)$$

to be consistent with designations of parameters, which we already used in Eq. 20-22. Deduced equation for the upper branch is:

$$\frac{\Delta C(T_c)}{\gamma \cdot T_c} = 1.38 \cdot \left(1 + 6.35 \cdot \left(\frac{k_B \cdot T_c}{\hbar \cdot \omega_{ln}}\right)\right) \qquad (28)$$

Details can be found in Fig. 7

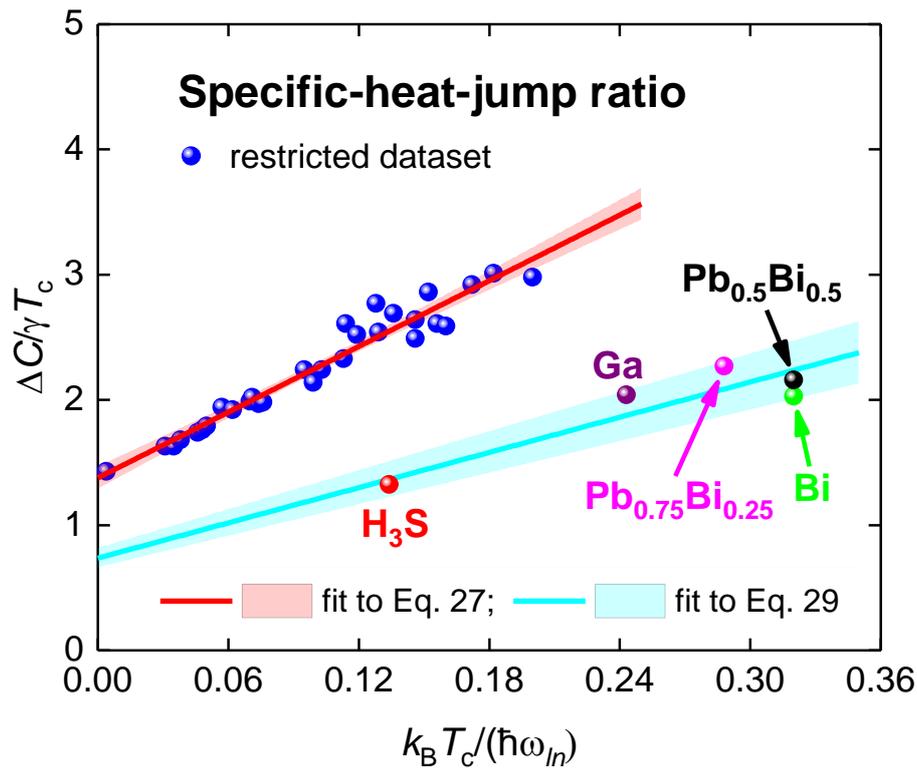

**Figure 7.** Specific-heat-jump ratio $\frac{\Delta C(T_c)}{\gamma \cdot T_c}$ vs $\frac{k_B \cdot T_c}{\hbar \cdot \omega_{ln}}$. The data is taken from Table X of Ref. 20 and data for $H_3S$ is taken from Refs. 25,26. Fits to Eq. 27 (blue data points, red curve) and Eq. 29 (cyan curve) are shown together with 95% confidence bands. The upper curve (Eq. 27) has the fitting parameter $A = 1.38 \pm 0.04$, $B = 6.35 \pm 0.47$, $R = 0.945$, mutual parameter dependence 0.928. The lower curve (Eq. 29) has the fitting parameter $A = 0.74 \pm 0.03$ and $R = 0.799$.

For the lower branch suggested fitting equation is similar to Eq. 23:

$$\frac{\Delta C(T_c)}{\gamma \cdot T_c} = A \cdot \left(1 + 6.35 \cdot \left(\frac{k_B \cdot T_c}{\hbar \cdot \omega_{ln}}\right)\right) \tag{29}$$

where $A = 0.74 \pm 0.03$, details can be found in Fig. 7.

### 3.6. Double-valued $\frac{2 \cdot \Delta(0)}{k_B \cdot T_c}$ vs $\frac{\Delta C(T_c)}{\gamma \cdot T_c}$ correction function

Because both ratios of $\frac{2 \cdot \Delta(0)}{k_B \cdot T_c}$ and of $\frac{\Delta C(T_c)}{\gamma \cdot T_c}$ are increasing/decreasing with the increase/decrease in the coupling strength, there is an expectation that these ratios are linked with each other by simple (in the first approximation, linear) relation. This question is discussed by J. P. Carbotte in Section V.E [20]. However, as this showed in Fig. 57 [20] and acknowledged in the text [20], that four superconductors with large $\frac{k_B \cdot T_c}{\hbar \cdot \omega_{ln}}$ values (which were excluded from the consideration in the rest of Ref. 20, and which we indicate by red circle/oval in Figs. 1,5) ruin simple relation of $\frac{2 \cdot \Delta(0)}{k_B \cdot T_c}$ vs $\frac{\Delta C(T_c)}{\gamma \cdot T_c}$.

However, we find that $\frac{2 \cdot \Delta(0)}{k_B \cdot T_c}$ vs $\frac{\Delta C(T_c)}{\gamma \cdot T_c}$ dependence is nicely matched the idea of double-valued correction functions, as we show in Fig. 8, where data for H$_3$S is added.

We fit the restricted dataset to simple linear function:

$$\frac{2 \cdot \Delta(0)}{k_B \cdot T_c} = E + F \cdot \frac{\Delta C(T_c)}{\gamma \cdot T_c} \tag{30}$$

where for the consistency with Eqs. 20-23,27-29 (which all utilize $\frac{k_B \cdot T_c}{\hbar \cdot \omega_{ln}}$ as primary variable), we use different designations for parameters, i.e. E and F in Eq. 30, instead of A, B, C in Eqs. 20-23,27-29.

The dataset for superconductors with large $\frac{k_B \cdot T_c}{\hbar \cdot \omega_{ln}}$ values and H$_3$S can be also fitted to Eq. 30, but because of small number of data points, 95% confidence band is reasonably large. The width of the confidence band has been reduced (Fig. 8), by fitting the dataset to function:

$$\frac{2 \cdot \Delta(0)}{k_B \cdot T_c} = E \cdot \left(1 + E \cdot \frac{\Delta C(T_c)}{\gamma \cdot T_c}\right) \qquad (31)$$

Deduced parameter is $E = 1.31 \pm 0.01$. This is important to note that all four superconductors with large $\frac{k_B \cdot T_c}{\hbar \cdot \omega_{ln}}$ values and H$_3$S fall into this more narrow 95% confidence band (Fig. 8).

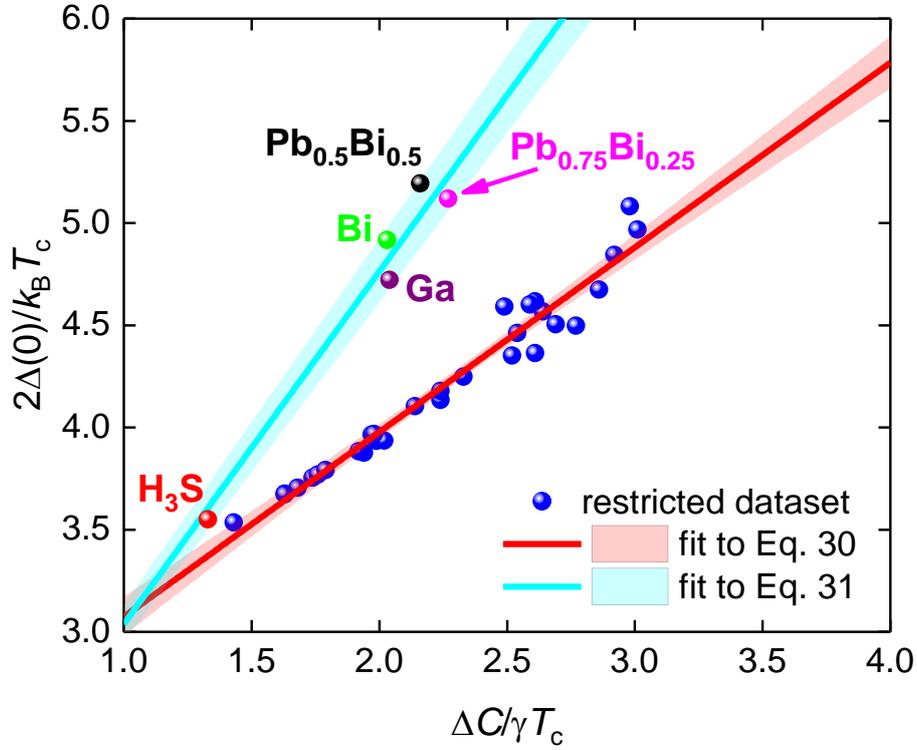

**Figure 8.** Full dataset for $\frac{2 \cdot \Delta(0)}{k_B \cdot T_c}$ vs $\frac{\Delta C(T_c)}{\gamma \cdot T_c}$ from Table IV, Table X and Fig. 57 in Ref. 20, and data point for highly-compressed H$_3$S [25,26]. Fits to Eqs. 30,31 are shown together with 95% confidence bands. Free-fitting parameters for the fit of restricted dataset to Eq. 30 (red curve) are: $E = 2.17 \pm 0.08$, $F = 0.90 \pm 0.03$, $R = 0.961$, mutual parameters dependence is 0.962. Free fitting parameter $E = 1.31 \pm 0.01$ and goodness of fit $R = 0.966$ are revealed from the fit to Eq. 31 the dataset for superconductors with large $\frac{k_B \cdot T_c}{\hbar \cdot \omega_{ln}}$ values and H$_3$S (cyan curve).

## IV. Conclusions

In this paper we analysed data for the gap-to-critical-temperature ratio, $\frac{2\cdot\Delta(0)}{k_B\cdot T_C}$, and for the specific-heat-jump ratio, $\frac{\Delta C(T_C)}{\gamma\cdot T_C}$, and find that:

1. Strong-coupling correction functions are double-valued functions.

2. All functions are nearly linear, and the upper branch of this function is described by the equation:

$\frac{2\cdot\Delta(0)}{k_B\cdot T_C} = 3.53 \cdot \left(1 + 3.53 \cdot \left(\frac{k_B\cdot T_C}{\hbar\cdot\omega_{ln}}\right)^{1.29}\right)$ the upper branch;

$\frac{2\cdot\Delta(0)}{k_B\cdot T_C} = 2.87 \cdot \left(1 + 3.53 \cdot \left(\frac{k_B\cdot T_C}{\hbar\cdot\omega_{ln}}\right)^{1.29}\right)$ the lower branch;

$\frac{\Delta C(T_C)}{\gamma\cdot T_C} = 1.38 \cdot \left(1 + 6.35 \cdot \left(\frac{k_B\cdot T_C}{\hbar\cdot\omega_{ln}}\right)\right)$ the upper branch;

$\frac{\Delta C(T_C)}{\gamma\cdot T_C} = 0.74 \cdot \left(1 + 6.35 \cdot \left(\frac{k_B\cdot T_C}{\hbar\cdot\omega_{ln}}\right)\right)$ the lower branch;

$\frac{2\cdot\Delta(0)}{k_B\cdot T_C} = 2.17 \cdot \left(1 + 0.90 \cdot \left(\frac{\Delta C(T_C)}{\gamma\cdot T_C}\right)\right)$ the upper branch;

$\frac{2\cdot\Delta(0)}{k_B\cdot T_C} = 1.31 \cdot \left(1 + 1.31 \cdot \left(\frac{\Delta C(T_C)}{\gamma\cdot T_C}\right)\right)$ the lower branch.

3. Highly-compressed $H_3S$ falls into lower branches, and this is the reason why at relatively large value of $\frac{k_B\cdot T_C}{\hbar\cdot\omega_{ln}} = 0.134$, this superconductor has the gap ratio, $\frac{2\cdot\Delta(0)}{k_B\cdot T_C}$, and specific-heat-jump ratio, $\frac{\Delta C(T_C)}{\gamma\cdot T_C}$, within weak-coupling limit of BCS theory.

**Acknowledgement**

Author thanks financial support provided by the state assignment of Minobrnauki of Russia (theme "Pressure" No. AAAA-A18-118020190104-3) and by Act 211 Government of the Russian Federation, contract No. 02.A03.21.0006.